\documentclass[11pt]{article}
\usepackage{graphicx} 
\usepackage[final]{epsfig}
\usepackage{amssymb,latexsym}
\usepackage{url}
\usepackage{amsmath}
\usepackage{ctable}
\usepackage{multirow}

\newcounter{mnotecount}[section]
\newcolumntype{C}{>{\small}c}
\newcolumntype{P}{>{\small}p}
\newcolumntype{M}{>{\small}m}

\begin{document}

\title{On light propagation in Swiss-Cheese cosmologies}
\author{
Sebastian J. Szybka\\
Astronomical Observatory, Jagellonian University, Krak\'ow\\
}
\date{}
\maketitle
\begin{abstract}
We study the effect of inhomogeneities on light propagation. The Sachs equations are solved numerically in the Swiss-Cheese models with inhomogeneities modelled by the \mbox{Lema\^itre-Tolman} solutions. Our results imply that, within the models we study, inhomogeneities may partially mimic the accelerated expansion of the Universe provided the light propagates through regions with lower than the average density. The effect of inhomogeneities is small and full randomization of the photons' trajectories reduces it to an insignificant level. 
\end{abstract}

\section{Introduction}

The  observations of type Ia supernovae interpreted within isotropic and homogeneous cosmological models imply the accelerated expansion of the Universe. 
These observations \cite{Riess:1998cb,Perlmutter:1998np} were made in late nineties and ignited a ``megabit bomb''\footnote{Stanis\l aw Lem, {\it Summa technologiae}, 1964.} of an enormous number of publications. Among many hypothesis that were proposed, the most conservative one assumed that inhomogeneities in the energy distribution may mimic the accelerated expansion. Although such a hypothesis does not solve the old cosmological constant problem (why this constant is so small), it gives a hope for understanding why the vacuum energy density in the concordance model is of the same order as the present matter energy density \cite{rasanen-2004-0402}. 

Inhomogeneities may have a twofold effect. Firstly, the averaging procedure in general relativity is not well understood yet. Hence, assuming homogeneity and then solving the Einstein equations could not lead to the proper metric \cite{0264-9381-4-6-025}. Secondly, the light propagates differently in inhomogeneous spacetimes. This may modify the luminosity distance -- redshift relation that is crucial for an interpretation of the type Ia supernovae data.

In this article, we follow \cite{einstraus,1969ApJ...155...89K,2007PhRvD..76l3004M,btt,Biswas:2007gi,Vanderveld:2008vi,2009MNRAS.400.2185C,Valkenburg:2009iw} and study exact solutions to the Einstein equations, the so-called Swiss-Cheese (SC) models. In such models, inhomogeneities do not influence the global dynamics by construction. Therefore, the averaging problem will not be investigated here. The SC models provide convenient settings for studies of light propagation in inhomogeneous spacetimes.

The SC models are constructed out of the Einstein-de Sitter (EdS) solution with spherical regions removed. The Lema\^itre-Tolman (LT) solutions are matched to the spacetime in the excised regions.\footnote{For another possibility see \cite{PhysRevD.82.103510}.} Since inhomogeneities in the real Universe are not spherically symmetric, it is not obvious how to choose density profiles of the inhomogeneous regions. Therefore, we treat the SC model as a toy model of the Universe and we search for a reasonable ``extremal'' setting to determine the maximal effect of inhomogeneities on the luminosity distance -- redshift curve. If shear is neglected, the upper bound on the luminosity distance for a given redshift is determined by the so-called {\it empty beam} formula \cite{1973ApJ...180L..31D}.\footnote{For light bundles which have not passed through a caustic \cite{1992grle.book.....S}.}

The numerical code that we have developed give us large freedom in the construction of models. The light may travel non-radially trough arbitrary size inhomogeneous regions whose centres do not have to lie in a plane or on a regular lattice. This allows us to investigate more general settings than these presented so far in the literature. We solve numerically the fully relativistic system of equations.

\section{Model}\label{sec:setting}

New exact solutions to the Einstein equations may be constructed out of the old ones with a help of a gluing technique. In this article, we consider inhomogeneous cosmological models that are made of the LT solutions. The large scale evolution is given by the EdS solution that belongs to the LT class. Therefore, we start with a short description of the LT solutions and matching conditions within this class.

The LT solutions are spherically symmetric solutions of the Einstein equations with a dust source 
\cite{lemaitre,tolman,bondi}. The corresponding line element
takes the following form in comoving coordinates
\begin{equation}\label{eq:LT}
ds^2=dt^2-\frac{R_{,r}^{\;2}}{1+2E}dr^2-R^2(d\theta^2+\sin^2\theta d\phi^2).
\end{equation}
$E=E(r)$ is an arbitrary function and \mbox{$R=R(t,r)$} satisfies equations
\begin{eqnarray}
\label{eq:E1}
R_{,t}^2&=&2E+\frac{2M}{R} -\frac 1 3 \Lambda R^2\;,\\
\label{eq:E2}
{8\pi}\epsilon &=& \frac{2M_{,r}}{R^2R_{,r}} \;,
\end{eqnarray}
where $M=M(r)$ is one more arbitrary function, $\Lambda$ is the cosmological constant and $\epsilon=\epsilon(t,r)$ is a dust
mass density. The solutions of \eqref{eq:E1} are given up to a third arbitrary
function $t_B(r)$. 
Moreover, \eqref{eq:LT}, \eqref{eq:E2} are covariant under
the coordinate transformation $r=f(r')$. In order to obtain a particular
solution one has to specify three functions $E(r)$, $M(r)$ 
and $t_B(r)$. Also other possibilities exist, e.g.\ in this article we will set $E(r)$, $\epsilon(t_{init},r)$ and $R(t_{init},r)$ at some $t_{init}$. All these functions have a simple physical interpretation \cite{bondi,plebkras}.
In principle one may specify them freely, but a general choice
will lead to pathologies. Supplementary conditions can assure regularity at
the centre and exclude shell-crossing singularities \cite{plebkras}.

For $\Lambda=0$, the solutions of \eqref{eq:E1} exist in an explicit form \cite{plebkras}.\\
When $E(r)<0$
\begin{eqnarray}\nonumber
R(t,r)&=&-\frac{M}{2E}(1-\cos\eta)\;,\\
\eta-\sin\eta&=&\frac{(-2E)^{3/2}}{M}(t-t_B(r))\;.\label{eq:sol1}
\end{eqnarray}
If $E(r)=0$, then 
\begin{equation}\label{eq:sol2}
R(t,r)=\left(\frac{9}{2}M(r)(t-t_B(r))^2\right)^{1/3}\;.
\end{equation}
When $E(r)>0$
\begin{eqnarray}\nonumber
R(t,r)&=&\frac{M}{2E}(\cosh\eta-1)\;,\\
\sinh\eta-\eta&=&\frac{(2E)^{3/2}}{M}(t-t_B(r))\;.\label{eq:sol3}
\end{eqnarray}

Two LT solutions can be joined
smoothly on the spherical hypersurface $\Sigma$ given by $r=r_b$. The Darmois junction conditions \cite{darmois}
state that the first fundamental
forms (intrinsic metrics) and the second fundamental forms (extrinsic
curvatures) calculated in terms of the coordinates on $\Sigma$ should be the
same on both sides of $\Sigma$. One may show that the matching conditions reduce to
\begin{equation}\label{eq:darmois}
\begin{array}{ccc}
E_1|_{\Sigma}=E_2|_{\Sigma}\;,&M_1|_{\Sigma}=M_2|_{\Sigma}\;,&(t_B)_1|_{\Sigma}=(t_B)_2|_{\Sigma}\;,
\end{array}
\end{equation}
where indices $1$, $2$ number matched solutions. The evolution outside the matching surface $\Sigma$ does not depend on the evolution inside $\Sigma$. The matching may be repeated arbitrary number of times as long as different $\Sigma$'s do not overlap. 

In this article, we study the EdS solution  with spherical regions excised. Inside each excised region the non-homogeneous LT solution is matched. Since the EdS spacetime is homogeneous, 
the matching surfaces do not have to be centred at the same point, but they may be scattered ``like holes in a cheese''. This construction leads to the non-homogeneous model of the Universe that is called the Swiss-Cheese (SC) model. 

\subsection{Setting}
Hereafter, we assume $\Lambda=0$. The large scale evolution is given by the LT solution with 
\begin{equation}\label{eq:EdS}
\begin{array}{ccc}
E(r)=0\;,& \epsilon(t,r)=\frac{1}{6\pi t^2}\;, & t_B(r)=0\;.
\end{array}
\end{equation} 
This choice corresponds to the EdS model. At some time $t_{init}$ we match inhomogeneities. They are modelled by the following LT solution. We  make the same choice as in \cite{2007PhRvD..76l3004M} and take
\begin{equation}\label{eq:ep}
\epsilon(t_{init},r)=A\exp^{-\frac{1}{2}\left(\frac{r-r_M}{\varsigma}\right)^2}+\delta\;.
\end{equation} 
Moreover, we assume $R(t_{init},r)=r$ and this together with \eqref{eq:sol3} determines $t_B(r)$. The formula for $E(r)$ follows from the assumption that at $t=t_{init}$ the speed of the angular expansion of the inhomogeneous regions equals to the speed of the expansion of the homogeneous part. It is 
\begin{equation}\label{eq:EE}
E(r)=\frac{2}{9}\left(\frac{r}{t_{init}}\right)^2-\frac{M}{r}\;.
\end{equation} 
Summarizing, the profile of inhomogeneities at time $t_{init}$ is given in comoving coordinates by the size of the inhomogeneity $r_b$, the position of a peak of the density $r_M$, it's amplitude $A$ and the width $\varsigma$, and finally, the parameter $\delta$ that controls the density at the centres of the inhomogeneous regions. The parameters $A$, $r_M$ are not arbitrary, but they are chosen to satisfy the matching conditions \eqref{eq:darmois}. We have found 
\begin{equation}\label{eq:A}
A=\left(\frac{1}{6\pi t_{init}^2}-\delta\right)\exp^{\frac{1}{2}\left(\frac{r_b-r_M}{\varsigma}\right)^2}\;,
\end{equation}
where $r_M$ is determined by solving numerically the equation $M_1|_{\Sigma}=M_2|_{\Sigma}$. This equation may be written with a help of an error function as
\begin{eqnarray}\nonumber
&&e^{-\frac{r_b r_M}{\varsigma^2}}
(\delta-\frac{1}{6\pi t^2_{init}}) 
\left(-6 e^{\frac{r_b^2}{2 \varsigma^2}} r_M \varsigma^2+2 e^{\frac{r_b r_M}{\varsigma^2}} (r_b^3+3 (r_b+r_M) \varsigma^2)\right.\\
&&\left.-3 e^{\frac{r_b^2+r_M^2}{2 \sigma^2}} \sqrt{2 \pi} \sigma (r_M^2+\sigma^2) (Erf(\frac{r_b-r_M}{\sqrt{2} \sigma})+Erf(\frac{r_M}{\sqrt{2} \varsigma}))\right)=0\;.\label{eq:rM}
\end{eqnarray}
$M_1|_{\Sigma}$ was found by integration of \eqref{eq:E2} with an assumption $M(0)=0$ and $M_2|_{\Sigma}=\frac{2}{9}\frac{r^3}{t_{init}^2}$ is a standard formula for the EdS spacetime.  

We cover the spacetime with many spherically symmetric comoving coordinate patches of the size $r_a$ centred at inhomogeneities. In principle, the free parameters $r_a$, $r_b$, $\varsigma$, $\delta$ may vary from one inhomogeneity to another. However, in order to reduce the number of degrees of freedom, we assume that for a given version of our model the ratio $r_a/r_b=s_a$ is fixed. The ratio $\varsigma/r_b=s_\varsigma$ and $\delta$ are chosen to be the same for all models studied in this article. We do not define the whole spacetime in advance, but construct it along the light beam (the congruence of null geodesics) that is evolved backward in time (from an observation to an emission). Therefore,
 for each inhomogeneous region there are two additional parameters. One of them, $\tilde{c}_\phi$ plays a role of an impact parameter of the light beam ($\tilde{c}_\phi$ is a normalized\footnote{At each entry to the new coordinate patch $c_\phi=(1+z)R(t,r_a)\tilde{c}_\phi$, where $c_\phi$ will be defined in Section \ref{sec:ge}.}
impact parameter, such that $\tilde{c}_\phi=1$ corresponds to the maximum value of the impact parameter). 
The second parameter is a phase of shear ($\arg\sigma$) of the beam at matching surfaces. A change of $\arg\sigma$ correspond to a rotation of the principal axes of shear in Sachs basis. If $\arg\sigma$ is not continuous, then centres of inhomogeneous regions may not lie on a two dimensional surface.
The both parameters $\tilde{c}_\phi$, $\arg\sigma$ are crucial, together with $r_b$, $s_a$, for the relative positions of inhomogeneous regions in spacetime. 
Finally, it is necessary to specify a present moment $t_0$ (or equivalently $H_0=\frac{2}{3t_0}$ --- the Hubble constant outside inhomogeneities at the present moment). 

The choice of numerical values of our free parameters is not completely arbitrary. We would like to avoid shell crossings in inhomogeneous regions and we do not want to have focal points for the congruence of null geodesics that we are going to study. In addition, the set of parameters that we choose should not be in obvious contradiction to observational data. The allowed values of parameters may be determined by the trial and error method. However, in order to be able to compare our results to \cite{2007PhRvD..76l3004M} we make similar choice and the LT inhomogeneous solutions describe central large underdensities surrounded by overdense shells. Of course, there are many other much different ``good'' settings.

Let us start with choosing units such that the numerical value of the dust density at the time of the observation $t_0$ equals to $1$. Thus, we set without loss of generality\footnote{In the EdS model $\rho=\frac{1}{6\pi t^2}$.} $t_0=1/\sqrt{6\pi}$. We assume that $t_{init}=0.2 t_0$. 
Next, we set $s_\varsigma=0.1$, $\delta=0.0025$ and the matching gives fixed ratio $r_M/r_b=0.881258$. In our work, we consider two values of the $s_a$ parameter: $s_a=1$ or $s_a=1.19048$. 

Let $N$ denote the number of inhomogeneous regions. There remain $3N$ free parameters ($(r_b)_{i=1\dots N}$, $(\tilde{c}_\phi)_{i=1\dots N}$, $(\arg\sigma)_{i=1\dots N}$) in the model. In order to reduce this number, we fix size of inhomogeneous regions $(r_b)_i$ for a given version of the model and assume that impact parameters $(\tilde{c}_\phi)_{i=1\dots N}$ are distributed in the interval $[-\tilde{c}_\phi^*,\tilde{c}_\phi^*]$ with the probability distribution function\footnote{The so-called ``non-aligned'' version of our SC models. Inhomogeneities are randomly spread in the spacetime and the probability that the beam will enter inhomogeneous region with a particular impact parameter $\tilde{c}_\phi$ is proportional to the circumference $2\pi\tilde{c}_\phi$. Such the probability distribution functions provides proper randomization of the path of the beam.} $|\tilde{c}_\phi/\tilde{c}_\phi^*|$ or with the uniform probability density.\footnote{The so-called ``aligned'' version of our SC models. Inhomogeneities are randomly spread in the plane in which the beam propagates. The beam is randomized only in the plane.} Moreover, the phase of shear $(\arg\sigma)_{i=1\dots N}$ is equal to zero or uniformly distributed in the interval
$(-\pi,\pi]$. We assumed that in each coordinate patch the beam should enter the inhomogeneous region. Hence, 
\begin{equation}\label{eq:ipr}
\tilde{c}_\phi^*=\min\{r_b/r_a,2r_b/r_a\sqrt{1-(r_b/r_a)^2}\}\;,
\end{equation} 
where the second value prevents inhomogeneities from overlapping and $r_b/r_a=1/s_a$. In our setting, we have $\tilde{c}_\phi^*=r_b/r_a\approx 0.84$.

In short, our models were defined in this Section. The differences between particular settings studied in this article are presented in Appendix \ref{appendixA}.

\section{Luminosity distance and angular diameter distance}\label{sec:lum}

The corrected luminosity distance \cite{1992grle.book.....S} from the source that is moving with the four-velocity $u^\alpha_S$ to the observer with $u^\alpha_O$ is defined by
\begin{equation}\label{eq:dlc}
\hat{d}_L=\left(\frac{\delta A_O}{\delta\Omega_S}\right)^\frac{1}{2},
\end{equation}
where $\delta A_O$ is the area of a cross-section of the beam seen by the observer (it does not depend on $u^\alpha_O$) 
and $\delta\Omega_S$ is
the solid angle of the beam that is measured at the source in a tangent 3-space orthogonal to $u^\alpha_S$ ($\delta\Omega_S$ depends on
$u^\alpha_S$
). Therefore, $\hat{d}_L$ depends only on the
position of the source, the position of the observer and the four-velocity of the source. Similarly, interchanging roles of the source
and the observer one can define the angular diameter distance $d_A=(\delta A_S/\delta\Omega_O)^{1/2}$. This quantity is not the observer independent
($\delta\Omega_O$ 
depends on $u^\alpha_O)$. The reciprocity theorem \cite{1992grle.book.....S} proves that in any spacetime
$\hat{d}_L=(1+z)d_A$, where $z$ is a
redshift of the source seen by the observer. 
The uncorrected luminosity distance is given by $d_L=(1+z)\hat{d}_L$. The additional redshift factor corresponds to the change of the energy of photons. Thus, $d_L$ has physical meaning and may be used to calculate the apparent brightness.  
Now, we may write
\begin{equation}\label{eq:dlm}
d_L=(1+z)\left(\frac{\delta A_O}{\delta\Omega_S}\right)^\frac{1}{2}\;,
\end{equation}
or using the reciprocity theorem
\begin{equation}\label{eq:dl}
d_L=(1+z)^2\left(\frac{\delta A_S}{\delta\Omega_O}\right)^\frac{1}{2}\;.
\end{equation}
In this article, we assume that the observer see the source within some small solid angle $\delta\Omega_O$ and we
trace back the beam to the source, evolve $A$ along the path and calculate $d_A$.  Finally, one may use \eqref{eq:dl} and the geodesic equation to
determine $d_L$ as a function of the redshift $z$. However, we have decided to present results in terms of $d_A$.

The evolution of the area of the cross-section of the beam is given by the Sachs optical equation \cite{sachs}
\begin{eqnarray}\label{eq:sachs}
\frac{d^2}{d\lambda^2}\sqrt{A}+\left(|\sigma|^2+\frac{1}{2}R_{\mu\nu}k^\mu k^\nu\right)\sqrt A=0\;,
\end{eqnarray} 
where $\lambda$ is an affine parameter, $k^\mu$ is a wave vector of the beam, $\sigma$ is complex shear
\begin{equation}\label{eq:shear}
|\sigma|^2=\frac{1}{2}(\nabla_\alpha k_\beta)( \nabla^\alpha k^\beta)-\theta^2\;,
\end{equation}
and $\theta$ is an expansion
\begin{equation}\label{eq:exp}
\theta=\frac{1}{2}\nabla_\mu k^\mu\;.
\end{equation}
We are interested in the evolution backward in time, so the initial conditions we adopt are
\begin{eqnarray}\label{eq:ic_Sachs}
\sqrt{A}|_O&=&0\;,\\
\nonumber \left.\frac{d\sqrt{A}}{d\lambda}\right|_O&=&-\sqrt{\delta\Omega_O}\;,
\end{eqnarray}
where the last equation follows from the definition of a solid angle. The observation corresponds to $\lambda=0$. 
The angular diameter distance does not depend on $\delta\Omega_O$ as long as $\delta\Omega_O$ is
small enough.

One procedure for solving \eqref{eq:sachs} would be to define initial conditions for a
congruence of null geodesics (in agreement with \eqref{eq:ic_Sachs}), solve the geodesic equation \eqref{eq:geodesic_back} with
these initial conditions, calculate the expansion \eqref{eq:exp}, shear \eqref{eq:shear} and finally solve
\eqref{eq:sachs} with initial conditions given by \eqref{eq:ic_Sachs}. However, it is more convenient to adopt a different
approach. 
Namely, one solves the geodesic equation \eqref{eq:geodesic_back} for a single central null geodesic and calculates shear and the expansion of the congruence
from the remaining Sachs optical equations\footnote{It is sufficient to use the equation $\theta=dA/d\lambda/(2A)$ with \eqref{eq:sachs3} supplemented by the assumption $(\theta|\sigma|)|_{\lambda=0}=0$.}
\begin{eqnarray}
\label{eq:sachs2}\frac{d\theta}{d\lambda}+\theta^2+|\sigma|^2&=&-\frac{1}{2}R_{\mu\nu}k^\mu k^\nu\;,\\
\label{eq:sachs3}\frac{d\sigma}{d\lambda}+2\theta\sigma&=&C_{\alpha\beta\gamma\delta}L_1^\alpha k^\beta k^\gamma L_1^\delta\;,
\end{eqnarray}
where $L_1$ is the spacelike vector orthogonal to the light ray (one of the vectors of the Sachs basis), $|\sigma|_{\lambda=0}=0$ and we set the phase of $\sigma$ to $0$ at $\lambda=0$. In our case, \eqref{eq:sachs3}
took a particularly simple form because of spherical symmetry of the LT solutions. It can be verified that 
\begin{equation}
L_1=\frac{1}{R}\frac{\partial}{\partial\theta}\;,
\end{equation} 
is a good choice of $L_1$ (see \cite{btt}).
In the parametrization \eqref{eq:LT} the right hand sides of \eqref{eq:sachs2}, \eqref{eq:sachs3} take the form 
\begin{eqnarray}
-\frac{1}{2}R_{\mu\nu}k^\mu k^\nu&=&-4\pi(1+z)^2\epsilon\;,\\
C_{\alpha\beta\gamma\delta}L_1^\alpha k^\beta k^\gamma L_1^\delta&=&2\pi(\frac{c_\phi}{R})^2\left(\epsilon-\frac{3M}{4\pi R^3}\right)\;,
\end{eqnarray}
where $c_\phi$ corresponds to the value of the impact parameter in a given coordinate patch.

The Sachs equations depend on the wave vector of the central light ray in the beam, so it is necessary to find the ray trajectory. In the next subsection, we describe the geodesic equation.

In a general spacetime, solving the Sachs and the geodesics equations involves numerical calculations. For spacetimes that are ``on-average--RW'' models, the effective procedure was suggested \cite{zeldovich64,1973ApJ...180L..31D}. This procedure neglects shear and for an ``on-average--EdS'' model gives the formula \cite{1992grle.book.....S}\footnote{This is the effective formula and no rigorous derivation is known at present.}
\begin{equation}\label{eq:emb}
d_A(z)=\frac{1}{H_0}\frac{(1+z)^\beta-(1+z)^{-\beta}}{2\beta(1+z)^\frac{5}{4}}\;,
\end{equation}
where $\beta=\frac{1}{4}\sqrt{25-24\tilde\alpha}$ and $\tilde\alpha$ is a dimensionless {\it smoothness parameter} (a mass-fraction of the matter in the Universe that is not bounded in galaxies). This is the so-called {\it partially filled beam} approximation and for $\tilde\alpha=0$ it is known as the {\it empty beam} approximation. We will compare \eqref{eq:emb} to our numerical results in Subection \ref{sec:nrPF}.

\subsection{Geodesic equation}\label{sec:ge}

In order to determine null geodesics we will make use of Killing symmetries of
the spacetime \eqref{eq:LT}. It follows from the form of the line element
\eqref{eq:LT} that one of Killing vectors implied by spherical symmetry
takes the form $\eta^\mu=(\partial/\partial \phi)^\mu$. Let $u^\mu$ be the four-velocity of a photon and let $\lambda$ be an affine parameter along photon trajectory
\begin{equation}\label{eq:u}
u=\frac{dt}{d\lambda}\left(\frac{\partial}{\partial t}\right)+
\frac{dr}{d\lambda}\left(\frac{\partial}{\partial r}\right)+
\frac{d\theta}{d\lambda}\left(\frac{\partial}{\partial \theta}\right)+
\frac{d\phi}{d\lambda}\left(\frac{\partial}{\partial \phi}\right)\;.
\end{equation}
It satisfies the geodesic equation $u^\mu\nabla_\mu u^\nu=0$.
The quantity $u^\mu \eta_\mu=d\phi/d\lambda R^2\sin^2\theta=c_\phi$ is conserved along
the photon path. We can easily choose our coordinate system in such a way that
at some point $\theta(\lambda_P)=0$. This implies that $c_\phi=0$ along
the whole trajectory and $\theta(\lambda)=0$ or
$\phi(\lambda)=const$. Therefore, we assume without loss of generality that one of
the angular variables is constant along the photon path. We choose our
coordinate system to have $\theta=\pi/2$. Thus,
$d\phi/d\lambda=c_\phi/R^2$. Moreover, $u^\mu$ is a null vector and this gives us the 
first integral of motion
$u^\mu u_\mu=0$. Using these equations and $t$-component of the
geodesic equation we obtain the system of the first
order differential equations
\begin{eqnarray}
\label{eq:geodesic_back}
\frac{dt}{d\lambda}&=&{z+1}\;,\\
\nonumber
\frac{dr}{d\lambda}&=&\pm\frac{1}{R_{,r}}\sqrt{(1+2E)\left((z+1)^{2}-\frac{c_\phi^2}{R^2}\right)}\;,\\
\nonumber
\frac{d\theta}{d\lambda}&=&0\;,\\
\nonumber
\frac{d\phi}{d\lambda}&=&\frac{c_\phi}{R^2}\;,\\
\nonumber
\frac{dz}{d\lambda}&=&-\frac{R_{,rt}}{R_{,r}}(1+z)^2+\frac{c_\phi^2}{R^2}\left(\frac{R_{,rt}}{R_{,r}}-\frac{R_{,t}}{R}\right)\;,
\end{eqnarray}
where $z(\lambda)$ is a new auxiliary function chosen up to an additive
constant. The second equation is inconvenient for numerical calculations because the expression under the square root can be a source of numerical problems for near zero values. Moreover, the plus and the minus sign 
can be encoded in initial conditions. Therefore,  we use the second order equation for $r(\lambda)$. It has the form
\begin{eqnarray}
\nonumber&&\frac{d^2r}{d\lambda^2}+2\frac{R_{,rt}(1+z)}{R_{,r}}\frac{dr}{d\lambda}
+\frac{1}{R_{,r}^2}\left((1+2E)\frac{R_{,rr}}{R_{,r}}-E_{,r}\right)\left((z+1)^2
-\frac{c_\phi^2}{R^2}\right)\\
&&-(1+2E)\frac{c_\phi^2}{R_{,r}R^3}=0\;.\label{eq:klk_back}
\end{eqnarray}

Let $u^\mu_S$ be the four-velocity of the source that is comoving with a dust. 
Since there are no
off-diagonal terms in the metric \eqref{eq:LT}, we have $1+z(\lambda)=u_{\mu}
u^\mu_S=\omega(\lambda)$, where $\omega(\lambda)$ is the frequency of a light signal of wave vector $u^\mu$ measured by the observer comoving with the source and a dust. If $z(0)=0$, then $z(\lambda)$ can be
interpreted as the redshift of the signal that was emitted at some $\lambda<0$ (by the source comoving with a dust) and was measured by the dust comoving observer at $\lambda=0$.

In this article, we assumed that the observer is at the matching surface between the EdS solution and the inhomogeneous region, hence we set $r(0)=r_a$. We have $dr/d\lambda\leq 0$ at $\lambda=0$ and the remaining initial conditions are: $t(0)=t_0$, $\theta(0)=\frac{\pi}{2}$, $\phi(0)=0$, $z(0)=0$.

\section{Numerical results}\label{sec:nr}

As a starting point we take the model with five inhomogeneous regions that was investigated by Marra, Kolb, Matarrese, Riotto (MKMR) in \cite{2007PhRvD..76l3004M}. 
It was shown there that the MKMR model almost reproduce the angular diameter distance -- redshift curve of the Robertson-Walker (RW) model with $\Omega_M=0.6$ and $\Omega_{\Lambda}=0.4$. This results suggest that the effect of inhomogeneities may be significant. We would like to verify how it depends on the assumptions made and the particular setting used by MKMR. Therefore, we generalize the MKMR model step by step and follow the changes in the angular diameter distance -- redshift curve. The similar analysis was done in \cite{Vanderveld:2008vi}, where the large effect in the MKMR model was explained as a result of insufficient randomization. The main difference between our study and \cite{Vanderveld:2008vi} comes from the fact that in \cite{Vanderveld:2008vi} the weak field gravitational lensing theory was used and shear was neglected. We solve numerically the fully relativistic system of equations and evaluate directly the effect of shear.

The differences between particular settings studied in this Section were summarized in Appendix \ref{appendixA}. The order of the keys in the figures corresponds to the order of the curves for the redshift $z=1.8$.

\vspace{-0.1cm}
\subsection{MKMR model}\label{subsec:gaps}
\vspace{-0.015cm}

In our setting the MKMR model corresponds to $s_\varsigma=0.1$, $s_a=1$, $r_b=0.042$ and $\arg\sigma=\tilde{c}_\phi=0$ in each inhomogeneous region. We reproduced the result of \cite{2007PhRvD..76l3004M}. The change in the angular diameter distance, $\Delta d_A(z)=d_A(z)-d_A(z)|_{RW}$, compared to the RW model with $\Omega_M=0.6$ and $\Omega_\Lambda=0.4$ is presented in Fig.\ \ref{fig:1}. It coincides with Fig.\ $13$ in \cite{2007PhRvD..76l3004M}.

\vspace{-0.1cm}
\subsection{Non-radial beams}
\vspace{-0.015cm}

In the MKMR model, the light propagates radially in each coordinate patch. Since the mass is concentrated in spherical shells and the density in the central part of the inhomogeneity is tiny, the beam travels through regions with lower density than the average. The angular diameter distance in such model should deviate from the EdS value as it already follows from the weak lensing analysis. It seems interesting to calculate the angular diameter distance for more typical beams, i.e.\ the beams that are not so much statistically distinguished. Such analysis requires the study of non-radial beams and randomization of the sequence of impact parameters. This extension of the MKMR is not a trivial one and leads to some subtle problems that will be addressed in this Subsection.

Firstly, for non-radial beams an impact parameter takes random values in each inhomogeneous region and centres of those regions are not lined up any more. Clearly, gaps between inhomogeneities and an upper limit for an impact parameter are necessary to avoid overlapping of inhomogeneities. Gaps are introduced by setting $s_a=1.19048$ which implies $r_a=0.05$. In addition, we assume that the absolute value of the impact parameter is not larger than $\tilde{c}_\phi^*=0.84$ (as defined in the equation \eqref{eq:ipr}). This assumption improves statistical properties of our models because it assures that the beam enters an inhomogeneous region in each coordinate patch.

Secondly, one has to decide how inhomogeneities are distributed in the spacetime. This is defined by the statistical properties of the sequence of impact parameters. We consider two possibilities in our paper. The spatial average density of the $t=const$ hypersurfaces does not deviate much from the EdS value. Hence, the natural assumption is that the average density along the beam which goes through inhomogeneous regions should not deviate from the average density along the beam that propagates in the EdS spacetime.\footnote{We acknowledge private communication with Syksy R\"as\"anen and Krzysztof Bo\-\mbox{lejko}. See also \cite{2011MNRAS.tmp...25B}.} To achieve this, impact parameters are distributed in $[-\tilde{c}_\phi^*,\tilde{c}_\phi^*]$ with the probability distribution function $|\tilde{c}_\phi/\tilde{c}_\phi^*|$. Such the probability distribution function corresponds to the trajectory of the typical beam that travels through inhomogeneous regions which are spread randomly in the spacetime.\footnote{The finite size of inhomogeneities and the assumption that the beam should enter an inhomogeneous region in each coordinate patch imply that the distribution of centres of inhomogeneities in spacetime is not uniform.} The second possibility we consider is that centres of all inhomogeneous regions are spread randomly in a plane and that the beam propagates in this plane. We will call this version of our models ``aligned'' and we will refer to remaining our models as to ``non-aligned''. It is assumed that the typical beam in the plane with aligned inhomogeneities corresponds to the sequence of impact parameters uniformly distributed in the interval $[-\tilde{c}_\phi^*,\tilde{c}_\phi^*]$. In our setting, this implies that the light propagates through regions with lower average density than the spatial average density in the model. Such configuration of inhomogeneities is unnatural, but it is convenient to model the selection effect, i.e.\ the light from the sources that are observed is more likely to travel through low density regions. 
\begin{figure}[t!]
\begin{center}
\vspace{-1cm}
\includegraphics[scale=0.48,angle=-90]{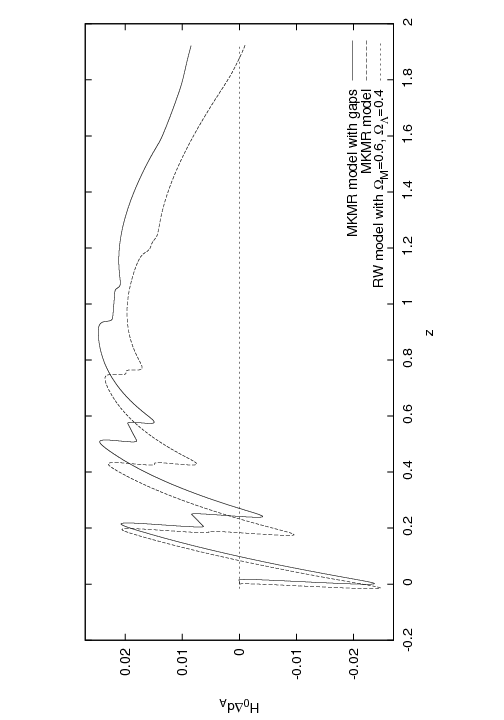}
\vspace{-1.2cm}
\caption{\label{fig:1}
The difference of the angular diameter distances (times $H_0$) between the MKMR model, the radial MKMR model with gaps and the RW model with $\Omega_M=0.6$, $\Omega_{\Lambda}=0.4$, ($\Delta d_A(z)=d_A(z)-d_A(z)|_{RW}$).
}
\vspace{-0.6cm}
\end{center}
\end{figure}

Next, the problem that is sometimes not handled properly in the literature is the value of shear at the boundary between coordinate patches. In general, the principal axes of shear do not have to coincide with our choice of Sachs basis. The rotation of principal axes of shear is necessary in non-aligned models and may be realised by a random change of the phase of shear ($\arg\sigma$) with the uniform probability distribution in $(-\pi,\pi]$. In aligned models, one should assume that the phase of shear do not change between coordinate patches. For a sake of curiosity, we will also consider the third possibility: an aligned model (a vanishing phase of shear) with the impact parameters probability distribution function $|\tilde{c}_\phi/\tilde{c}_\phi^*|$.

In the remaining part of the paper, the following terminology will be used. Whenever we will refer to the aligned version of the non-radial SC model, we will explicitly indicate that. We will also explicitly indicate whenever only radial beams will be studied in the model (if different than the MKMR model). The remaining models are assumed to be non-aligned and to contain non-radial beams.

Let us start with applying the extension presented in this Subsection to the MKMR model. In the first step we add gaps.  
The angular diameter -- redshift curve in the radial MKMR model with gaps does not reveal big changes --- Fig.\ \ref{fig:1}. The inhomogeneities appear for a little bigger redshift, as expected.
\begin{figure}[t!]
\begin{center}
\vspace{-1.9cm}
\includegraphics[scale=0.40,angle=0]{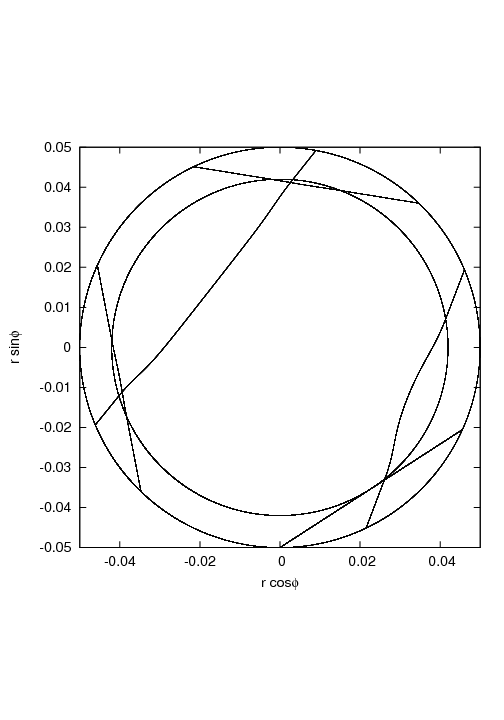}
\vspace{-1.8cm}
\caption{\label{fig:2}
The typical non-radial trajectories (comoving coordinates) in the MKMR model with gaps. The inner circle corresponds to $r=r_b$ and the outer to $r=r_a$. All inhomogeneous regions were overimposed on the single plot and the planes of the trajectories were rotated.}
\end{center}
\end{figure}
Next, we randomize impact parameters.
The non-radial beams in the MKMR model with gaps are curved. We present them in Fig.\ \ref{fig:2}, where $\tilde c_\phi$ is distributed in $[-\tilde{c}_\phi^*,\tilde{c}_\phi^*]$ with the probability density $|\tilde c_\phi/\tilde{c}_\phi^*|$ (non-aligned setting). The angular diameter distance is lower than in the EdS model because the beam spends ``more time'' in the region of higher density --- Fig.\ \ref{fig:3}. The effect of inhomogeneities on the angular diameter distance is opposite than for radial beams. However, this is not a typical property of non-radial beams, but an effect of week randomization (only five inhomogeneous regions).

\subsection{Small inhomogeneities}\label{subsec:smh}

\begin{figure}[t!]
\begin{center}
\vspace{-1.3cm}
 \includegraphics[scale=0.48,angle=-90]{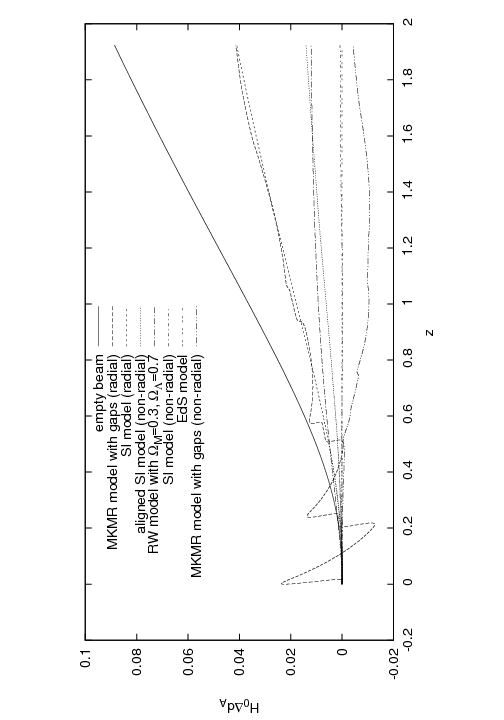}
\vspace{-1.1cm}
\caption{\label{fig:3}
The difference of the angular diameter distances (times $H_0$) between the {\it empty beam} approximation, the SI models, the MKMR models with gaps (radial and non-radial), the RW model with $\Omega_M=0.3$, $\Omega_\Lambda=0.7$ and the EdS model ($\Delta d_A(z)=d_A(z)-d_A(z)|_{EdS}$).
$H_0\Delta d_A$ for the RW model was multiplied by a factor $10^{-1}$ to make the comparison more explicit. The order of the keys corresponds to the order of the curves for the redshift $z=1.8$.
}

\vspace{-0.6cm}
\end{center}
\end{figure}

\begin{figure}[t!]
\begin{center}
\vspace{-1.3cm}
\includegraphics[scale=0.48,angle=-90]{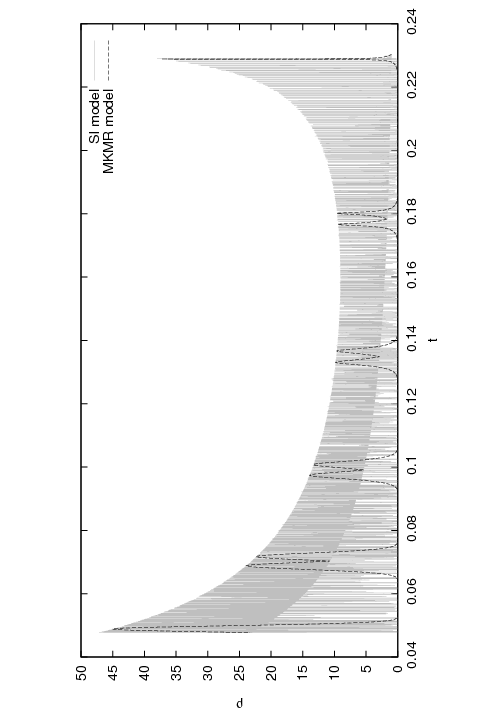}
\vspace{-1cm}
 \caption{\label{fig:4}
The density along the beam in the model with small inhomogeneous regions (the SI model) and in the model with five large inhomogeneous regions (the MKMR model).
}
\vspace{-0.4cm}
\includegraphics[scale=0.48,angle=-90]{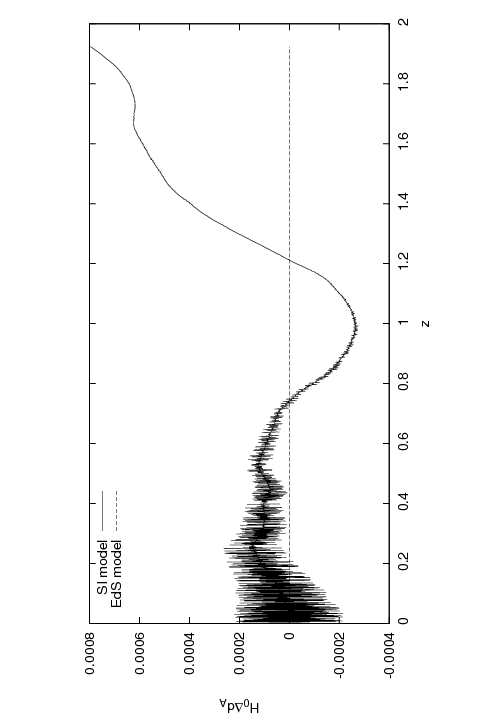}
\vspace{-1cm}
\caption{\label{fig:5}
The difference of the angular diameter distances (times $H_0$) between the SI model and the EdS model, ($\Delta d_A(z)=d_A(z)-d_A(z)|_{EdS}$).
}
\end{center}
\end{figure}

In the MKMR model with gaps (non-radial), the angular diameter distance $d_A(z)$ depends strongly on the sequence of impact parameters. One should average $d_A(z)$ over many runs to obtain a reliable result (like in \cite{Vanderveld:2008vi}) or reduce the size of inhomogeneities to obtain the better statistic in a single run.
We decided to reduce the size of inhomogeneities $100$ times (we set $r_b=0.00042$).\footnote{The diameter of the inhomogeneous region is around $10$ Mpc at the moment of an observation.} In the model with small inhomogeneous regions (hereafter, the SI model), the density along the beam is a fast varying function --- Fig.\ \ref{fig:4}. For radial beams, the angular diameter distance does not change much, but in a general case the effect of inhomogeneities on the angular diameter distance is negligible --- Figs \ref{fig:3}, \ref{fig:5}. For non-radial beams in the aligned SI model,
the effect of inhomogeneities on light propagation for $z<1.5$ is lower than $10\%$ of what is needed to explain the accelerated expansion without introducing the cosmological constant. It is around one-third of the effect in the original MKMR model. 
Since this is the maximal effect we have observed in our models (after randomization of the trajectory of the beam), the aligned SI model corresponds to our ``extremal'' setting.
The {\it empty beam} formula \eqref{eq:emb} gives the largest angular diameter distance.

\begin{figure}[t!]
\begin{center}
 \includegraphics[scale=0.48,angle=0]{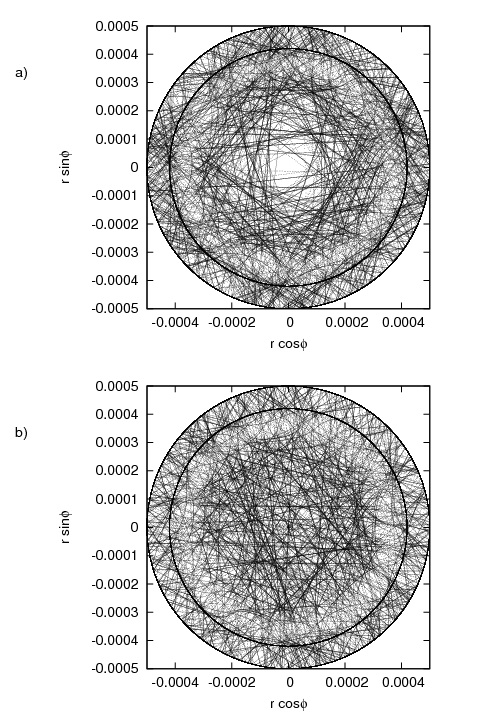}
 \caption{\label{fig:6}
The typical non-radial trajectories (comoving coordinates) in a) the SI model, b) the aligned SI model. The inner circle corresponds to $r=r_b$ and the outer to $r=r_a$. The planes of trajectories were rotated and all inhomogeneous regions were overimposed on the single plot.}
\end{center}
\end{figure}

The typical trajectories in the SI model (non-aligned and aligned) are presented in Fig.\ \ref{fig:6}. It follows from Figs \ref{fig:3}, \ref{fig:5} that the effect of inhomogeneities on the angular diameter distance is reduced in our models to the insignificant level by proper randomization of the beam's trajectory.

\clearpage

\subsection{Shear}\label{subsec:sh}

In the previous subsections, we have introduced non-radial beams with non-vanishing shear along the trajectories. 
In most models studied in the literature (including  \cite{Vanderveld:2008vi}), shear is assumed to be negligible for the angular diameter distance -- redshift relation. We have verified that this is indeed true in our models.
\begin{figure}[tp!]
\begin{center}
\vspace{-1.5cm}
\includegraphics[scale=0.48,angle=-90]{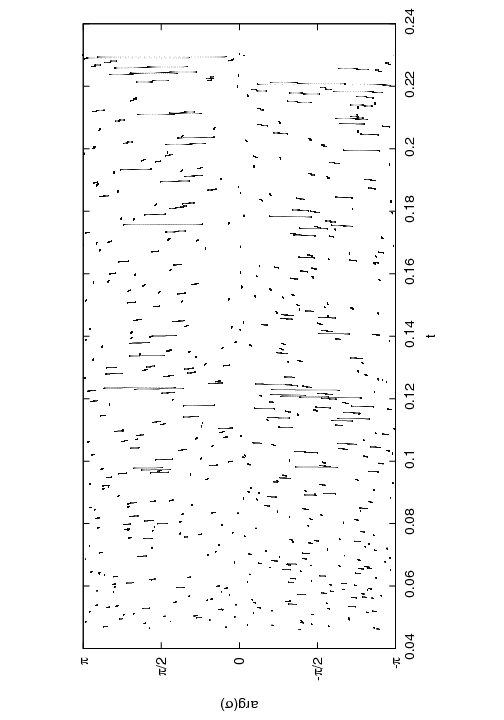}
\vspace{-1cm}
\caption{\label{fig:7}
The angle between the Sachs basis and the principal axes of shear in the SI model.}
\vspace{-0.0cm}
 \includegraphics[scale=0.48,angle=-90]{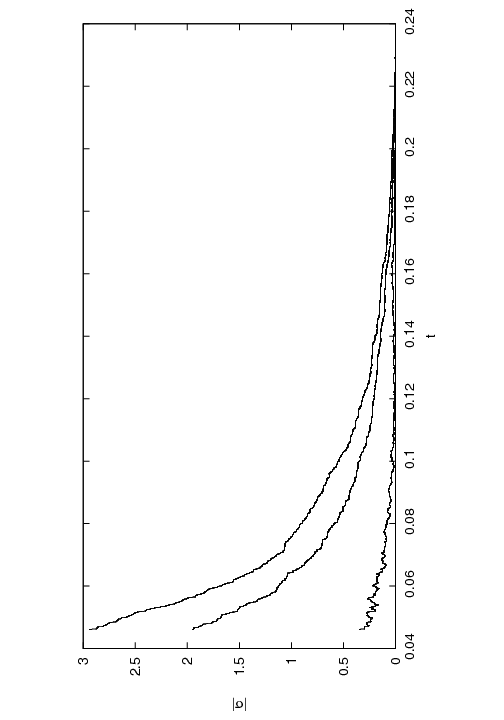}
\vspace{-1cm}
 \caption{\label{fig:8}
The effect of averaging of shear by inhomogeneities whose centres are not aligned. The bottom curve corresponds to the SI model. The middle curve corresponds to the aligned SI model. The top curve corresponds to the aligned model with the probability distribution function of the impact parameters $|\tilde{c}_\phi/\tilde{c}_\phi^*|$ (the aligned SI* model).}
\end{center}
\end{figure}
\begin{figure}[th!]
\begin{center}
\vspace{-1.3cm}
 \includegraphics[scale=0.48,angle=-90]{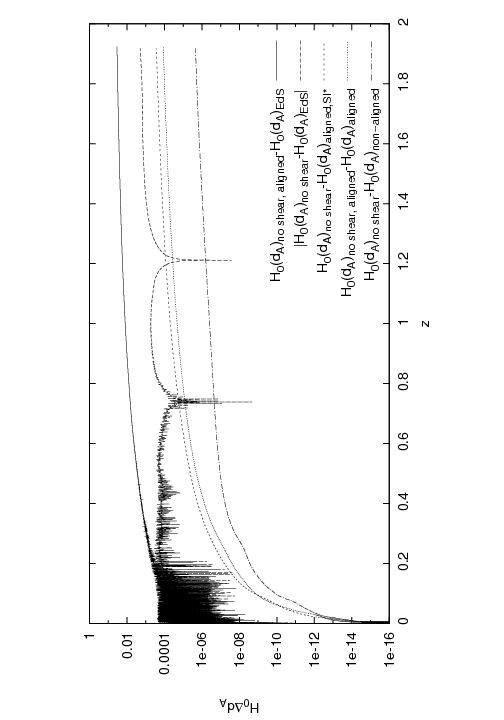}
\end{center}
\vspace{-1.2cm}
 \caption{\label{fig:9}
The difference of the angular diameter distances (times $H_0$) between the SI model with neglected shear (aligned, non-aligned), the EdS model, the aligned and the non-aligned SI model, and the aligned SI* model. The order of the keys corresponds to the order of the curves for large redshifts.
}
\vspace{-0.4cm}
\end{figure}

In the non-aligned models, 
the angle between Sachs basis and principal axes of shear is uncorrelated between inhomogeneous regions (Fig.\ \ref{fig:7}). In contrast to that, in the aligned models the phase of shear is constant and equal to zero.

We compare the value of shear in three versions of the SI models: non-aligned, aligned, and aligned (a vanishing phase of shear) with the probability distribution function of the impact parameters $|\tilde{c}_\phi/\tilde{c}_\phi^*|$. The last model was introduced to verify validity of some calculations presented in the literature. It will be distinguished with a symbol SI*. 
The modulus of shear is a smooth function and is presented in Fig.\ \ref{fig:8}. It is interesting to observe in this figure the averaging effect of non-aligned inhomogeneities on shear.

Let $n$ be a number of inhomogeneous regions encountered by the beam. At first sight the non-aligned positions
of inhomogeneities (in the non-aligned SI model) lead to a random walk in $|\sigma|\sim \sqrt{n}$ in contrast to the aligned case where $|\sigma|\sim n$ (the aligned SI model). We have verified that the curves in Fig.\ \ref{fig:8} do not satisfy this relations. Since structures are growing in time, the expected value of the change of shear between neighbouring inhomogeneities is time dependent. Therefore, this process may be studied as a continuous time random walk.

In Fig.\ \ref{fig:9} we present the difference of the angular diameter distances between the SI model, the aligned SI model, the aligned SI* model and the SI model with neglected shear (non-aligned and aligned), and the EdS model. This comparison reveals that the effect of shear for non-aligned SI models is tiny (for $z=1.5$ it less than $0.1\%$ of the correction to the EdS angular diameter distance). For aligned SI models, it is less than $1\%$ (for $z=1.5$). However, the aligned SI* model (with the probability distribution function $|\tilde{c}_\phi/\tilde{c}_\phi^*|$), overestimate it around two orders of magnitude (for $z=1.5$ the effect of shear is around $10\%$ of the correction to the EdS angular diameter distance). 

The random number generator in our code was initialized with the same random seed. Therefore, three bottom curves in Fig.\ \ref{fig:9} are not jagged for small redshifts.

\begin{figure}[t!]
\begin{center}
\vspace{-1.4cm}
 \includegraphics[scale=0.48,angle=-90]{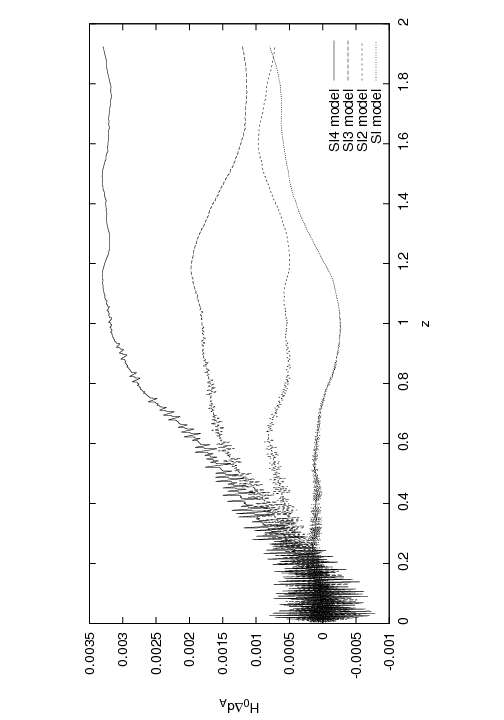}
\end{center}
\vspace{-1cm}
 \caption{\label{fig:10}
The difference of the angular diameter distances (times $H_0$) in four models with different sizes of inhomogeneous regions SI, SI2, SI3, SI4 and the EdS model ($\Delta d_A(z)=d_A(z)-d_A(z)_{EdS}$). The order of the keys corresponds to the order of the curves for the redshift $z=1.8$.}
\end{figure}

\subsection{Partially filled beam approximation and different sizes of inhomogeneities}\label{sec:nrPF}

It was argued in \cite{Vanderveld:2008vi} that the dimming of supernovae in the MKMR model may be roughly estimated with a help of the {\it partially filled beam} approximation \eqref{eq:emb}. We apply this approximation to five models: the SI model, three SIn models with inhomogeneities n-times bigger than in the SI model (where $n=2,3,4$), and the MKMR model. The angular diameter distance, in variations of the SI model, seems to depend slightly on the size of inhomogeneities (Fig.\ \ref{fig:10}) and the sequence of impact parameters induces big statistical fluctuations.\footnote{The code was initialized with the same random seed, but sizes of inhomogeneities are different. Therefore, the different parts of a random sequence corresponds to different redshifts.} 
\begin{figure}[t!]
\begin{center}
\vspace{-1.2cm}
 \includegraphics[scale=0.48,angle=-90]{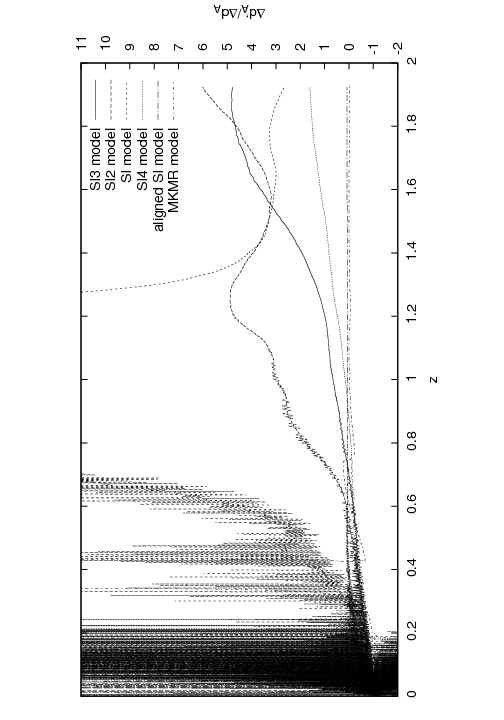}
\end{center}
\vspace{-1cm}
 \caption{\label{fig:11}
The {\it partially filled beam} formula compared to the numerically calculated angular diameter distance for the SI, SI2, SI3, SI4, MKMR models, ($\Delta d_A'(z)=d_A(z)_{partially\;filled\;beam}-d_A(z)$, $\Delta d_A(z)=d_A(z)-(d_A(z))_{EdS}$).
The smoothness parameter was calculated for each model using \eqref{eq:alpha}.
The order of the keys corresponds to the order of the curves for the redshift $z=1.8$ and $z=1.7$.}
\end{figure}
We suppose that it is not size of inhomogeneities that matters, but randomization, i.e.\ how the average density along the beam deviates from the EdS value.\footnote{The radial versions of MKMR and SI models give similar angular diameter distance as may be seen in Fig.\ \ref{fig:3}.}
This may be seen as follows. Let us calculate the smoothness parameter $\tilde\alpha$ for each model. The smoothness parameter is a mass-fraction of the matter in the Universe that is not bounded in galaxies. We calculate it for our SC models using 
\begin{equation}\label{eq:alpha}
\tilde\alpha=\frac{1}{\lambda_0-\lambda_S}\int_{\lambda_0}^{\lambda_S}\frac{\rho_{SC}(\lambda)}{\rho_{EdS}(t_{SC}(\lambda))}d\lambda\;,
\end{equation}
where $\lambda_0=0$ denotes an affine parameter at the observer and  $\lambda_S$ corresponds to the emission. In our models, the spatial average density of the $t=const$ hypersurfaces is approximately equal to the density in the EdS model. Therefore, one may expect that $\tilde\alpha$ should not diverge far from $1$. However, the bigger are inhomogeneities, the weaker is randomization of the path of the beam. Since low density regions occupy larger volume in the studied models, it is more likely that $\tilde\alpha$ will decrease with increasing size of inhomogeneities (the variance will grow). 
We have found $\tilde\alpha$ to be approximately equal to $0.964$, $0.939$, $0.915$, $0.895$, $0.815$, $0.412$ for the SI, SI2, SI3, SI4, aligned SI, MKMR models, respectively.  
For SI models that contain hundreds of inhomogeneities, $\tilde\alpha$ is slightly lower than $1$. The values of $\tilde\alpha$ for particular models together with Fig.\ \ref{fig:10} suggest that the effect of inhomogeneities on the angular diameter distance depends on the average density along the beam (as expected from the week lensing approximation).

We plotted Fig.\ \ref{fig:11} which shows the difference between the {\it partially filled beam} formula and the numerical result for the approximated model. It reveals that the {\it partially filled beam} approximation is acceptable for smaller values of $\tilde\alpha$. The accuracy is around $1\%$ of the correction to the EdS angular diameter distance for the MKMR model and $8\%$ for the aligned SI model (for $z=1.5$). In our models, the {\it partially filled beam} approximation does not work for larger values\footnote{For $\tilde\alpha\simeq 1$, the effect of inhomogeneities seems to be smaller than statistical fluctuations.} of $\tilde\alpha$.
\section{Summary}

In this article, the effect of inhomogeneities on light propagation was investigated in the framework of the SC models. We have examined the angular diameter distance -- redshift relation. This type of distance is related to the luminosity distance by the reciprocity theorem. Therefore, the theoretical angular diameter distance -- redshift relation is crucial for an interpretation of the type Ia supernovae data.

Our analysis confirms that inhomogeneities may partially mimic the accelerated expansion of the Universe provided the light propagates through regions with lower than the average density. The effect is small and it becomes negligible if the average density along the beam does not deviate from the corresponding EdS value. In light of our work and the weak field gravitational lensing analysis \cite{Vanderveld:2008vi}, the result \cite{2007PhRvD..76l3004M} that suggest more significant influence of inhomogeneities is due to a peculiar setting of the underlying model. The precise size of the effect depends on the details of the model that was studied. Since the SC models are toy-models of the real Universe, it is speculative to base on them the final conclusions. Nevertheless, what our analysis shows is that, within the models we have studied, the effect of inhomogeneities remains too small\footnote{In the aligned SI model (our ``extremal'' setting with randomized photon trajectories), the effect of inhomogeneities on the angular diameter distance was one order smaller than what is needed to eliminate dark energy.}
to explain the type Ia supernovae observations without dark energy. Our analysis of the fully relativistic and non-linear models did not reveal any stronger effect on the angular diameter distance than that predicted by the {\it partially filled beam} approximation.  Randomization reduces the effect considerably in accordance with \cite{Vanderveld:2008vi}, \cite{2009MNRAS.400.2185C}, \cite{btt2} (and the others). 

We have directly evaluated the effect of shear on the angular diameter distance. Within our models, the effect of shear was negligible, but the models that do not take into account the rotation of the principal axes of shear (e.g.\ \cite{Valkenburg:2009iw}) may overestimate its effect around two orders of magnitude. In these models, the overestimated shear plays a minor role and it may lead to the small underestimation of the effect of inhomogeneities.

Our results suggest that the size of inhomogeneities is not crucial for the angular diameter distance, provided that non-radial models are sufficiently randomized. We have found that the {\it partially filled beam} formula \eqref{eq:emb} gives good approximation to the angular diameter distance if the average density along the beam is much lower than the corresponding EdS value.

Finally, we stress the analysis presented in this article does not touch directly the ``averaging problem'' in general relativity. The SC models behave on large scales by construction as the RW models, thus the influence of inhomogeneities on the global expansion rate of the Universe cannot be studied within this framework.

\vspace{0.5cm}

\noindent{\sc Acknowledgements}

The main part of this research was conducted during my stay at the University of Geneva and was founded by the Foundation for Polish Science through its {\sc Kolumb} program. I would like to express my gratitude to Ruth Durrer for hospitality and useful discussions. I thank Krzysztof Bolejko, Piotr Chru\'sciel, Bart\l{}omiej Kos, Syksy R\"as\"anen for comments and Zdzis\l{}aw Golda, Andrzej Woszczyna for suggesting this topic to me. I acknowledge private communication with Christina Sormani and Wessel Valkenburg.

The numerical calculations were carried out with the supercomputer ``Deszno'' purchased thanks
to the financial support of the European Regional Development Fund
in the framework of the Polish Innovation Economy Operational Program
(contract no.\ POIG.02.01.00-12-023/08). I acknowledge the use of {\sc GNU Scientific
Library} \cite{GSL} and {\sc Mathematica} together with the {\sc xAct} \cite{xAct} package.

\clearpage
\appendix

\section{Models and parameters}\label{appendixA}

%\scalebox{0.71}{
\ctable[pos=h!,
caption = {The settings of our models. The remaining parameters coincide for all models and were defined in Section \ref{sec:setting}. The parameters $\tilde{c}_\phi$, $\arg(\sigma)$ are randomly chosen at each entry to a new coordinate patch. If not indicated, the default probability distribution of the impact parameter $\tilde{c}_\phi$ is $|\tilde{c}_\phi/\tilde{c}_\phi^*|$ (in the interval $[-\tilde{c}_\phi^*,\tilde{c}_\phi^*]$), where $\tilde{c}_\phi^*=0.84$. The phase of shear $\arg(\sigma)$ is uniformly distributed in $(-\pi,\pi]$.},botcap
]
{M{4.2cm}M{1cm}M{1cm}M{1.3cm}M{1cm}M{1.4cm}}
{\tnote[$\dagger$]{Impact parameters were uniformly distributed in $[-\tilde{c}_\phi^*,\tilde{c}_\phi^*]$, where $\tilde{c}_\phi^*=0.84$.}}
{
\toprule[0.4mm]
MODEL & $r_b$ & $s_a$ & $\tilde{c}_\phi$ & $\arg(\sigma)$ & figures \\
\toprule[0.4mm]
\parbox{4.4cm}{MKMR\\ (radial, aligned)} & $0.042$ & $1$ & $0$ & $0$ & $1,4,11$ \\
\midrule
\parbox{4.4cm}{MKMR with gaps\\ (radial, aligned)} & $0.042$ & $1.19048$ & $0$ & $0$ & $1,3$ \\
\midrule
\parbox{4.4cm}{MKMR with gaps\\ (non-radial, non-aligned)} & $0.042$ & $1.19048$ & $[-\tilde{c}_\phi^*,\tilde{c}_\phi^*]$ & $(-\pi,\pi]$ & $2,3$ \\
\midrule
\parbox{4.4cm}{SI\\ (radial, aligned)} & $0.00042$ & $1.19048$ & $0$ & $0$ & $3$ \\
\midrule
\parbox{4.4cm}{SI\\ (non-radial, aligned)\tmark[$\dagger$]} & $0.00042$ & $1.19048$ & $[-\tilde{c}_\phi^*,\tilde{c}_\phi^*]$ & $0$ & $3,6,8,9,11$ \\
\midrule
\parbox{4.4cm}{SI\\ (non-radial, non-aligned)} & $0.00042$ & $1.19048$ & $[-\tilde{c}_\phi^*,\tilde{c}_\phi^*]$ & $(-\pi,\pi]$ & \parbox{1.4cm}{$3,4,5,6,7,$\\$8,9,10,11$} \\
\midrule
\parbox{4.4cm}{SI (non-radial, non-aligned,\\ shear neglected)} & $0.00042$ & $1.19048$ & $[-\tilde{c}_\phi^*,\tilde{c}_\phi^*]$ & --- & $9$ \\
\midrule
\parbox{4.4cm}{SI (non-radial, aligned,\\ shear neglected)\tmark[$\dagger$]} & $0.00042$ & $1.19048$ & $[-\tilde{c}_\phi^*,\tilde{c}_\phi^*]$ & --- & $9$ \\
\midrule
\parbox{4.4cm}{SI*\\ (non-radial, aligned)} & $0.00042$ & $1.19048$ & $[-\tilde{c}_\phi^*,\tilde{c}_\phi^*]$ & $ 0 $ & $8,9$ \\
\midrule
\parbox{4.4cm}{SI2\\ (non-radial, non-aligned)} & $0.00084$ & $1.19048$ & $[-\tilde{c}_\phi^*,\tilde{c}_\phi^*]$ & $(-\pi,\pi]$ & $10,11$ \\
\midrule
\parbox{4.4cm}{SI3\\ (non-radial, non-aligned)} & $0.00126$ & $1.19048$ & $[-\tilde{c}_\phi^*,\tilde{c}_\phi^*]$ & $(-\pi,\pi]$ & $10,11$ \\
\midrule
\parbox{4.4cm}{SI4\\ (non-radial, non-aligned)} & $0.00168$ & $1.19048$ & $[-\tilde{c}_\phi^*,\tilde{c}_\phi^*]$ & $(-\pi,\pi]$ & $10,11$ \\
\bottomrule[0.4mm]
}
\clearpage
\bibliographystyle{amsplain}  
\bibliography{sc.bib}
\end{document}